\documentclass[journal=acsccc,manuscript=article]{achemso}

\usepackage{chemformula} 
\usepackage[T1]{fontenc} 
\usepackage{graphicx}

\renewcommand{\vec}{\mathbf}

\renewcommand{\r}{\vec{r}}
\renewcommand{\k}{\vec{k}}

\newcommand{\ve}{\varepsilon}

\SectionNumbersOn

\author{Alexey V. Nenashev}
\affiliation{Faculty of Physics,
Philipps-Universit\"{a}t Marburg,
Marburg 35032, Germany}
\altaffiliation{On leave of absence from Rzhanov Institute of Semiconductor Physics and the Novosibirsk State University, Russia}
\email{nenashev_isp@mail.ru}

\author{Florian Gebhard}
\affiliation
{Faculty of Physics,
Philipps-Universit\"{a}t Marburg,
Marburg 35032, Germany}

\author{Klaus Meerholz}
\affiliation{Department f\"{u}r Chemie, Universit\"{a}t zu K\"{o}ln,
Greinstra\ss e 4-6, 50939 K\"{o}ln, Germany}

\author{Sergei D. Baranovskii}
\affiliation{Department f\"{u}r Chemie, Universit\"{a}t zu K\"{o}ln,
Greinstra\ss e 4-6, 50939 K\"{o}ln, Germany}
\alsoaffiliation{Faculty of Physics,
Philipps-Universit\"{a}t Marburg,
Marburg 35032, Germany}
\email{sergei.baranovski@physik.uni-marburg.de}

\title[An \textsf{achemso} demo]
  {Computation of the spatial distribution of charge-carrier density in disordered media}

\abbreviations{IR,NMR,UV}

\begin{document}







\begin{abstract}
The space- and temperature-dependent electron distribution $n(r,T)$ determines optoelectronic properties of disordered semiconductors. It is a challenging task to get access to $n(r,T)$ in random potentials, avoiding the time-consuming numerical solution of the Schr\"{o}dinger equation. We present several numerical techniques targeted to fulfill this task. For a degenerate system with Fermi statistics, a numerical approach based on a matrix inversion and that based on a system of linear equations are developed. For a non-degenerate system with Boltzmann statistics, a numerical technique based on a universal low-pass filter and one based on random wave functions are introduced. The high accuracy of the approximate calculations are checked by comparison with the exact quantum-mechanical solutions.

\includegraphics[width=\linewidth]{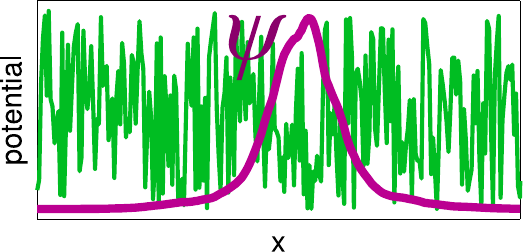}

\end{abstract}

\section{Introduction}
Disordered materials, such as amorphous organic and inorganic semiconductors and semiconductor alloys, play an important role in modern optoelectronics for computing, communications, photovoltaics, sensing, and light emission~\cite{Baranovskii2006Book,Masenda2021,Baranovskii2022Mini}. The spatial and energy disorder creates a random potential, which decisively affects electron states. Among other effects, a disorder potential causes the spatial localization of electrons in the low-energy range. The knowledge of the space- and temperature-dependent electron distribution $n(r,T)$, particularly in localized states created by random potentials, is required to understand charge transport and light absorption/emission in disordered semiconductors. The distribution $n(r,T)$ is most straightforwardly obtained by solving the Schr\"{o}dinger equation in the presence of a disorder potential. However, this procedure is extremely demanding with respect to computation facilities. It is hardly affordable for realistically large chemically complex systems. Therefore, it is highly desirable to develop theoretical tools to get access to $n(r,T)$ without solving the Schr\"{o}dinger equation.

One of the currently mostly used theoretical tools to reveal the individual features of localized states in a random potential is the so-called localization-landscape theory (LLT)~\cite{Arnold2016,LL1_2017,LL2_2017,LL3_2017}. In the LLT, the random potential is converted into some effective potential, which drastically simplifies all calculations. However, recent studies~\cite{Gebhard2023,Nenashev2023} revealed substantial problems of the LLT. For instance, the effective potential in the LLT lacks a temperature dependence that is necessary to describe $n(r,T)$ appropriately. Moreover, the LLT is seen to be equivalent to the Lorentzian filter applied to a random potential~\cite{Gebhard2023}. Such a choice of the filter function is rather unfortunate. The Lorentzian filter yields a significantly larger number of localized states in a random potential than the number of such states obtained via the exact solution of the Schr\"{o}dinger equation~\cite{Gebhard2023}.  Therefore, more developed computational techniques are desirable.

Here we develop two numerical techniques to reveal $n(r,T)$ in disordered systems under degenerate conditions controlled by Fermi statistics, avoiding the time-consuming numerical solution of the Schr\"{o}dinger equation. One of the techniques is based on converting the Hamiltonian into a matrix, which, being subjected to several multiplications with itself succeeded by inverting the outcome, yields the distribution $n(r,T)$. The other technique replaces the operation of matrix inversion by solving a system of linear equations controlled by the matrix generated from the Hamiltonian.

We also describe two recently developed computational techniques for calculations of $n(r,T)$ in non-degenerate systems controlled by Boltzmann statistics~\cite{Gebhard2023,Nenashev2023}.
One algorithm is based on applying a temperature-dependent universal low-pass filter (ULF) to the random potential $V(r)$. This yields a temperature-dependent effective potential, $W(r,T)$, that enables a quasiclassical calculation of particle density $n(r,T)$.
The ULF algorithm employs Fast-Fourier Transformation for calculating the effective potential $W$, enabling the analysis of very large systems.

The other algorithm is based on the recursive application of the Hamiltonian to multiple sets of random wave functions (RWF) for a specific realization of the random potential $V(r)$. Following the repeated application of the thermal operator, the temperature-dependent electron density is determined by averaging the outcomes over different RWF sets. This procedure offers several advantages over the widely-used LLT. Unlike the LLT, which relies on an adjustable parameter that can only be determined through comparison with the exact solution~\cite{Gebhard2023,Nenashev2023}, the RWF scheme lacks any adjustable parameters and works across all temperatures. Additionally, the accuracy of the RWF approach in computing $n(r,T)$ can systematically be improved, whereas the accuracy of the LLT is inherently limited.

\section{Calculation of $n(r,T)$ in a degenerate system controlled by Fermi statistics}
\label{sec:howto-degenerate}

To be definite, we consider a disorder potential acting on electrons, characterized by Gaussian statistics (`white noise'), i.e.,
the potential obeys $ \langle V(\vec{r}\,) \rangle_{\rm R} =0$
with the auto-correlation function~\cite{Halperin1966}
\begin{equation}
 \langle V(\vec{r}\,) V(\vec{r}\,{}^{\prime})
 \rangle_{\rm R} = S \delta\left(\vec{r}-\vec{r}\,{}^{\prime}\right)\; ,
 \label{eq:white_noise}
\end{equation}
where $\langle \ldots \rangle_{\rm R}$ indicates
the average over many realizations~${\rm R}$
of the random potential and $S$ is the strength of the interaction. This quantity $S$ yields natural definitions for the characteristic length scale and for the characteristic energy scale in the form
\begin{equation}
\ell_0=	\left( \frac{\hbar^4}{m^2 S} \right)^{1/(4-d)} \; ,
	\label{eq:length-unit}
\end{equation}
\begin{equation}
T_0	= \frac{1}{k_B} \left( \frac{m^d S^2}{\hbar^{2d}} \right)^{1/(4-d)} \; ,
	\label{eq:energy-unit}
\end{equation}
where $d$ is the space dimensionality and $k_B$ is the Boltzmann constant.

We consider here a collection of non-interacting electrons in some external potential in the thermodynamic equilibrium that is characterized by the temperature $T$ and the Fermi energy $\ve_f$. The goal is to develop an effective numerical method for the calculation of the electron density (concentration) $n(\mathbf{r},T)$ as a function of coordinates $\mathbf{r}$ in a degenerate system. The electron density can be defined as
\begin{equation}\label{eq:n-definition}
	n(\mathbf{r},T) = 2 \sum_a |\psi_a(\mathbf{r})|^2 f(\ve_a) .
\end{equation}
In this equation, summation index $a$ labels the electron eigenstates, i.~e., solutions of the Schr\"odonger equation $\hat H\psi_a = \ve_a\psi_a$, where $\psi_a(\mathbf{r})$ and $\ve_a$ are the wavefunction and the energy of the eigenstate; $f(\ve)$ is the Fermi function,
\begin{equation}\label{eq:f-definition}
	f(\ve) = \frac{1}{\exp[(\ve-\ve_f)/k_BT]+1} \, ,
\end{equation}
and factor 2 before the sum in Eq.~(\ref{eq:n-definition}) accounts for the two possible spin orientations.


Below we assume that the system under study is discretized with a finite-difference (or tight-binding) method. A wavefunction $\psi(\mathbf{r})$ is therefore represented as a collection of probability amplitudes $\psi_1, ..., \psi_L$ at $L$ points (grid nodes) evenly distributed in space. The Hamiltonian $\hat H$ is a matrix of size $L \times L$. Equation~(\ref{eq:n-definition}) for $n(r,T)$ in this discrete setting obtains the form
\begin{equation}\label{eq:n-discrete}
	n_i = \frac{2}{\Delta V} \sum_{a=1}^L |\psi_{a,i}|^2 f(\ve_a) ,
\end{equation}
where $n_i$ is the electron density at grid node $i\in\{1, ..., L\}$; $\psi_{a,i}$ is the value of eigenfunction $\psi_a$ at node $i$; $\ve_a$ is the electron energy that corresponds this eigenfunction; and $\Delta V$ is the spatial volume per one grid node (in the one-dimensional case, $\Delta V$ is simply the distance between nodes).

Equation~(\ref{eq:n-discrete}) represents a standard way for a numerical calculation of the electron density in degenerate systems. However this way requires the solution of the eigenvalue problem, which takes a large amount of computational resources for the large size $L$ of the Hamiltonian matrix. Below we suggest two methods that allow one to speed up the calculation of $n(r,T)$. In Method~1 (see Section~\ref{sec:inversion}), the numerical solution of the eigenvalue problem is replaced by a matrix inversion. The latter numerical task is much faster than the solution of the eigenvalue problem in the case of a \emph{sparse matrix}, in which almost all entries are equal to zero. In Method~2 (see Section~\ref{sec:linear-equations}), we employ a numerical solution of a system of linear equation, which is even faster than the matrix inversion. Performance of Methods~1 and~2, as compared to the standard method based on the eigenvalue problem, is tested in Section~\ref{sec:example} on an example of a one-dimensional disordered system with one occupied band.

The idea of Methods~1 and~2 is based on a simple observation that the shape of the function
\begin{equation}\label{eq:my-function}
	y(x) = \frac{1}{x^\mathcal N + 1} \, ,
\end{equation}
where a number $\mathcal N$ is large, resembles the shape of the Fermi function in the vicinity of point $x=1$. Other approaches for approximating the Fermi function are also possible.~\cite{Goedecker1999,Ceriotti2008,Lin2009prb,Bowler2012} Replacing $x$ with an appropriate linear function of the Hamiltonian is the essence of Methods~1 and~2. The detailed justification is given in Appendix~\ref{sec:why}.


\subsection{Method 1: matrix inversion}
\label{sec:inversion}

The method is based on the Hamiltonian $\hat H$ (a matrix $L \times L$), the Fermi energy $\ve_f$, and two additional parameters: a ``reference energy'' $\ve_0$ and the number of iterations $N$. These parameters are related to the temperature $T$,
\begin{equation}\label{eq:parameters-and-T}
	k_BT = \frac{|\ve_f-\ve_0|}{2^N} \, .
\end{equation}
Details for the choice of these parameters are given in Appendix~\ref{sec:choice}.

The method starts from composing the matrix $\hat A_0$ from the Hamiltonian
\begin{equation}\label{eq:make_A0}
	\hat A_0 = \frac{\hat H - \ve_0 \hat I}{\ve_f - \ve_0} \, ,
\end{equation}
where $\hat I$ is the $L \times L$ unit matrix. Then, the matrix $\hat A_0$ is squared $N$ times
\begin{equation}\label{eq:squaring}
	\hat A_p = \left(\hat A_{p-1}\right)^2, \quad p=1,2,\ldots,N.
\end{equation}
To the resulting matrix $\hat A_N$ we add the unit matrix and invert the sum to get a new matrix
\begin{equation}\label{eq:make-B}
	\hat B = \left( \hat A_N + \hat I \right)^{-1} .
\end{equation}
Finally, the electron density is obtained from the diagonal elements $B_{ii}$ of the matrix $\hat B$:
\begin{equation}\label{eq:n-by-inversion-1}
	n_i = \frac{2}{\Delta V} \, B_{ii} \quad \mathrm{if} \quad \ve_0 < \ve_f \, ,
\end{equation}
\begin{equation}\label{eq:n-by-inversion-2}
	n_i = \frac{2}{\Delta V} \, (1-B_{ii}) \quad \mathrm{if} \quad \ve_0 > \ve_f \, .
\end{equation}


\subsection{Method 2: solving a system of linear equation}
\label{sec:linear-equations}

\begin{figure}
\includegraphics[width=3cm]{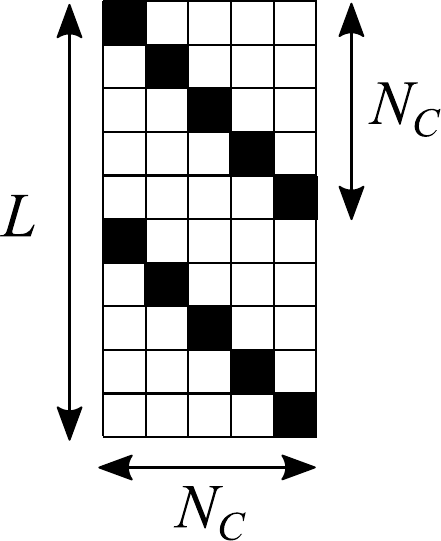}
\caption{Sketch of matrix $\hat U$ for the one-dimensional case. White and black squares represent zeros and unities, respectively.}
\label{fig:U}
\end{figure}

Similarly to Method~1, the input parameters are $\ve_f$, $\ve_0$ and $N$, which are related to the temperature $T$ by Eq.~(\ref{eq:parameters-and-T}). In addition, this method requires one more parameter $N_C$.
The choice of $N_C$ is discussed in Appendix~\ref{sec:choice}.
For a given $N_C$, we compose a matrix $\hat U$ of size $L \times N_C$ that obeys the following conditions (see Fig.~\ref{fig:U} for the shape of this matrix in the one-dimensional case):
\begin{itemize}
	\item each entry of matrix $\hat U$ is equal to either 0 or 1;
	\item in each row of matrix $\hat U$, exactly one entry is equal to 1;
	\item in each column of matrix $\hat U$, the nodes with nonzero entries are placed spatially as far from each other as possible. For example, in the one-dimensional case, the unities in each column are separated by $(N_C-1)$ zeros, as illustrated in Fig.~\ref{fig:U}.
\end{itemize}
The matrix $\hat A_N$ is calculated, as done in Method~1, see Eqs.~(\ref{eq:make_A0}) and~(\ref{eq:squaring}). However, in contrast to Method~1, no matrix inversion is necessary. Instead, a system of linear equations
\begin{equation}\label{eq:linear-equations}
	\left( \hat A_N + \hat I \right) \hat X = \hat U
\end{equation}
is to be solved with respect to an unknown matrix $\hat X$ of size $L \times N_C$. This is a computationally easier task in comparison to the matrix inversion.

Finally, the electron density is calculated as
\begin{equation}\label{eq:n-by-linear-equations-1}
	n_i = \frac{2}{\Delta V} \, \sum_{a=1}^{N_C} X_{ia} U_{ia} \quad \mathrm{if} \quad \ve_0 < \ve_f \, ,
\end{equation}
\begin{equation}\label{eq:n-by-linear-equations-2}
	n_i = \frac{2}{\Delta V} \, \left( 1 - \sum_{a=1}^{N_C} X_{ia} U_{ia} \right) \quad \mathrm{if} \quad \ve_0 > \ve_f \, .
\end{equation}


\subsection{Numerical example: a one-dimensional disordered system with one occupied band}
\label{sec:example}

\begin{figure}
\includegraphics[width=10cm]{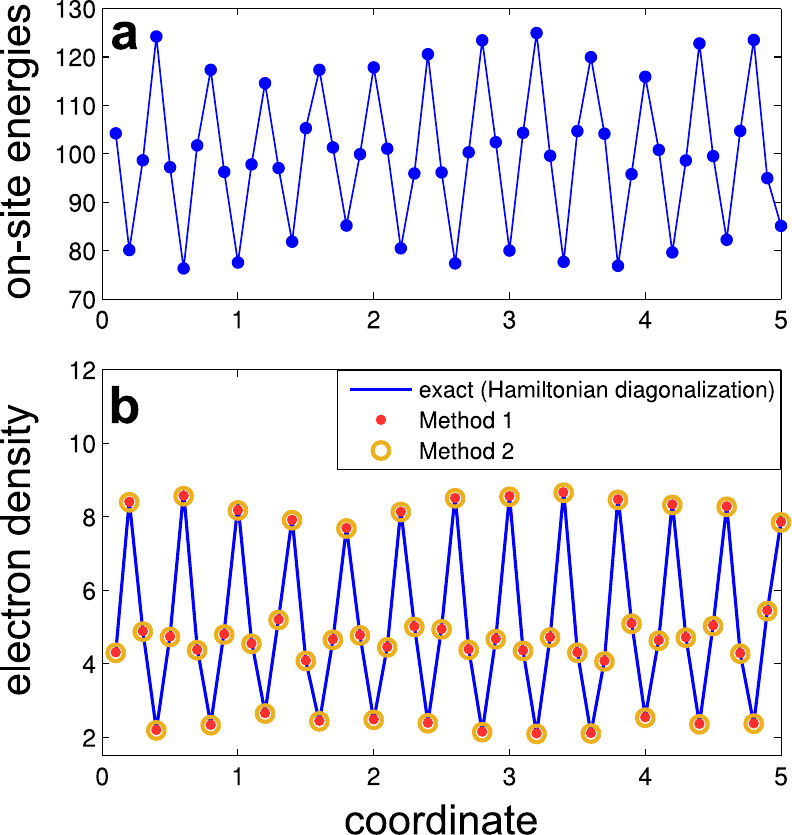}
\caption{Numerical example of one-dimensional tight-binding model: (a) on-site energies $H_{jj}$ at different nodes; (b) electron density in the lowest energy band calculated by three methods: exact diagonalization of the Hamiltonian (Eq.~(\ref{eq:n-discrete-filledVB}), blue lines), Method~1 (red dots), and Method~2 (orange circles).}
\label{fig:density-degenerate}
\end{figure}

Let us compare the performance of different methods to calculate $n(r,T)$ in a simple one-dimensional tight-binding model with energy bands and disorder. We consider a linear chain of $L$ lattice nodes at a distance $a=0.1$ from each other. We measure distances and energies in the units determined by Eqs.~(\ref{eq:length-unit}) and (\ref{eq:energy-unit}). In these units, the hopping integrals between neighboring nodes $H_{i,i+1}$ and $H_{i+1,i}$ are equal to $-1/(2a^2) = -50$. The on-site energies $H_{jj}$ are chosen to be
\begin{equation}\label{eq:on-site-energies}
	H_{jj} = V_0 + V_1 \cos(\pi j/2) + V_2 \xi_j \, ,
\end{equation}
where $\xi_j$ are random numbers uniformly distributed in the range $-1<\xi_j<1$. All other matrix elements of the Hamiltonian $\hat H$ are equal to zero. Periodic boundary conditions apply.
The constant term $V_0 = 100$ makes the lower boundary of the energy spectrum $\ve_{\mathrm{min}}$ to be near zero, and the higher boundary $\ve_{\mathrm{max}}$ to be $\approx 200$. The amplitude of periodic variations of the potential is $V_1 = 20$. The amplitude of the random-noise potential $V_2 = \sqrt{30}$ is chosen such that the ``disorder strength'' $S$ is equal to unity. As an example, we show one realization of the on-site energies in Fig.~\ref{fig:density-degenerate}a.

As a consequence of the periodic potential (the second term in the r.h.s. of Eq.~(\ref{eq:on-site-energies})), the electron energy spectrum consists of the four energy bands separated from each other by band gaps. We consider the situation when the lowest energy band is completely occupied by electrons, and three other bands are empty (quarter filling). The electron density in such a case is expressed as
\begin{equation}\label{eq:n-discrete-filledVB}
	n_i = \frac{2}{\Delta V} \sum_{a \in \mathrm{OB}} |\psi_{a,i}|^2 ,
\end{equation}
where summation is performed over the eigenstates of the occupied band.

The wave functions $\psi_a$, that enter Eq.~(\ref{eq:n-discrete-filledVB}), can be obtained by diagonalizing the Hamiltonian.
An example of the calculated electron density distribution is shown in Fig.~\ref{fig:density-degenerate}b by blue lines.

Also we show in Fig.~\ref{fig:density-degenerate}b the approximated electron density obtained by Method~1 (red dots) and by Method~2 (orange circles). The parameters for these methods are: $\ve_f = 28.5$ (the middle of the lowest band gap), $\ve_0 = 10$, $N=3$, and $N_C=30$. One can see that both Methods provide quite accurate results for the electron density.

\begin{figure}
\includegraphics[width=12cm]{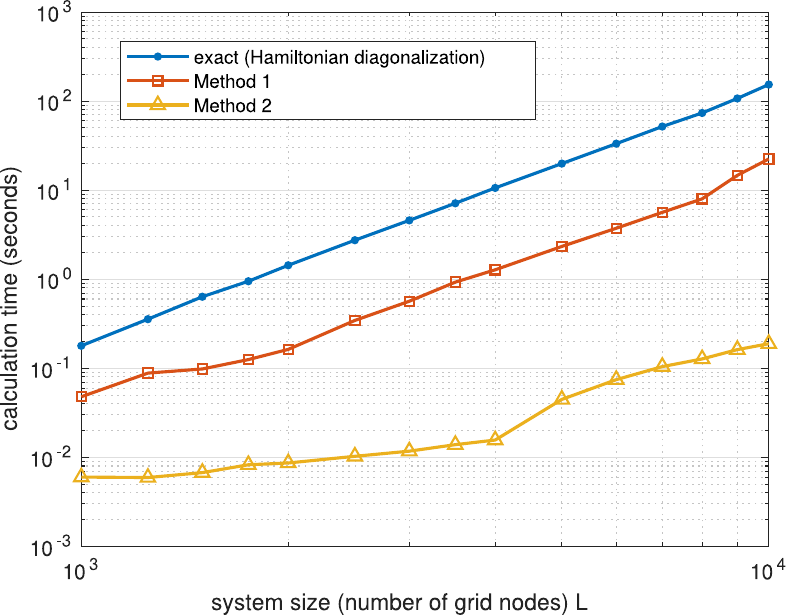}
\caption{The average calculation time for the electron density in the one-dimensional tight-binding model shown in Fig.~\ref{fig:density-degenerate} as a function of the number of nodes $L$. Different curves correspond to different numerical methods.}
\label{fig:timing}
\end{figure}

In Fig.~\ref{fig:timing} we compare the time required by three ways of calculating the electron density---the exact method based on the Hamiltonian diagonalization, Method~1, and Method~2---on a desktop PC with MATLAB used for the matrix manipulations. One can see that Method~1, which employs matrix inversion, is approximately one order of magnitude faster than the usual method of Hamiltonian diagonalization. Method~2 provides additional speedup by more than an order of magnitude.

Note that Method~1 can be further improved by using advanced techniques to calculate the diagonal part of the inverted matrix~\cite{Lin2009cms,Tang2012,Li2013}.  For the sake of simplicity, we use here the standard MATLAB functions in Method~1. Even in such a non-optimized setting, this Method demonstrates a substantial speedup in comparison with matrix diagonalization.

\section{Calculation of $n(r,T)$ in a non-degenerate system controlled by Boltzmann statistics}
\label{sec:LPF}

\subsection{Low-pass filter (LF) approach}
\label{sec:LF_approach}

\subsubsection{Motivation}
\label{sec:motivation}

\begin{figure}
  \includegraphics[width=10cm]{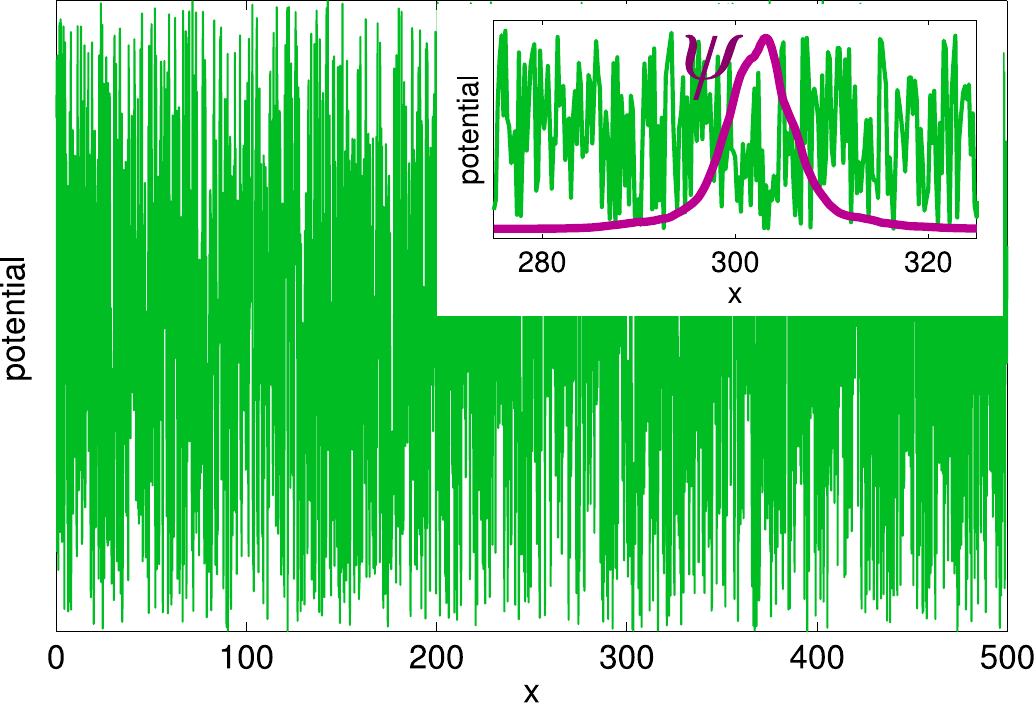}
  \caption{Realization for the white-noise disorder potential
on a one-dimensional strip. Insert: the wave function $\psi(x)$ of the state with the lowest energy in the region $275 \leq x \leq 325$. The coordinate $x$ is dimensionless in the units given by Eq.~(\ref{eq:length-unit}). }
 \label{fig:P1_fig1a}
\end{figure}

Already in the 1960s, Halperin and Lax recognized that electrons in a random potential cannot follow very short-range potential fluctuations.~\cite{Halperin1966} This effect is illustrated in Fig.~\ref{fig:P1_fig1a}, where the random white-noise potential $V(x)$ in one dimension is depicted by the green solid line. The detailed shape of $V(x)$ in the region $275 \leq x \leq 325$ (in the dimensionless units given by Eq.~(\ref{eq:length-unit})) is compared with the shape of the wave function $\psi$ for the lowest energy state in this spatial region. Apparently, the characteristic width $\ell_{wf}$ of the wave function, even for low-energy localized states, is substantially larger than the spatial scale of the fluctuations of the disorder potential $V(r)$. The latter scale in semiconductor alloys is of the order of the lattice constant $a \approx 0.5$~nm. The strong inequality $\ell_{wf} \gg a$ suggests that electrons in localized states are affected only by the mean disorder potential, averaged over the space scale $\ell_{wf}$.
It is, therefore, not necessary to solve exactly the Schr\"{o}dinger equation with the real disorder potential in order to get access to the individual features of electron states. Instead, one can apply to the disorder potential $V(r)$  a low-pass filter~\cite{Halperin1966} (LF) that smoothes the spatial fluctuations of $V(r)$.

Halperin and Lax suggested the square of the wave function as the filter function~\cite{Halperin1966}. The width of the filter function, $\ell \approx \ell_{wf}$ was adjusted dependent on the energy, $\varepsilon$, of the localized state, $\ell \propto |\varepsilon|^{-1/2}$, where $\varepsilon$ is counted from the band edge in the absence of disorder.
A variational approach was used to determine the shape of the low-energy density of states in a random potential~\cite{Halperin1966}. Baranovskii and Efros~\cite{Baranovskii1978} addressed the same problem by a slightly different variational approach and confirmed the result of Halperin and Lax.

Neither of the two groups considered, however, the individual features of localized states, being focused solely on the structure of the density of states in the low-energy region~\cite{Halperin1966,Baranovskii1978}. Our aim here is, in contrast, to calculate the spatial distribution of electron density, $n(r,T)$, in a given disorder potential $V(r)$. We start below with the definition of the low-pass filter and then formulate the algorithm for the calculation of $n(r,T)$.

\subsubsection{Definiton of a low-pass filter (LF)}
\label{sec:Definiton_low-pass_filter}

In one dimension, the low-pass filter (LF) is determined by the operation
\begin{equation}
  W(x) = \int {\rm d} x' \Gamma(x-x') V(x') \, ,
  \label{eq:convolution}
\end{equation}
where the filter function $\Gamma$ should contain the appropriate length scale $\ell$. This operation converts the real disorder potential $V(x)$ into the smooth effective potential $W(x)$. For instance, one can try a Lorentzian function $\Gamma^{\rm L}$,
\begin{equation}
  \Gamma^{\rm L}(x)
  =  \frac{e^{-|x|/\ell_{\rm L}}}{2\ell_{\rm L}}
  \; ,
  \label{eq:convolutionLorentz}
\end{equation}
because using as LF a Lorentzian function with $\ell_{\rm{L}} = 0.27 \, \ell_0$ has recently been proven~\cite{Gebhard2023} equivalent to the popular LLT approach~\cite{Arnold2016,LL1_2017,LL2_2017,LL3_2017}.

\begin{figure}
  \includegraphics[width=\textwidth]{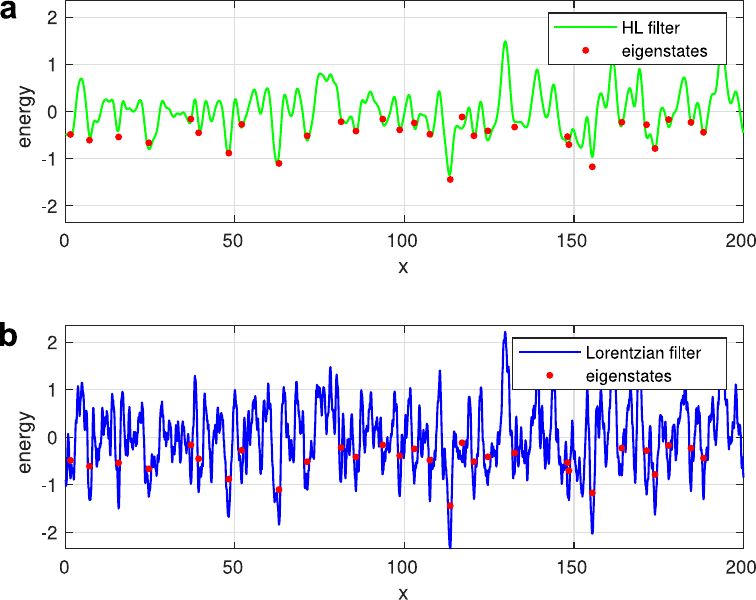}
  \caption{(a) Effective potential for a Halperin-Lax low-pass
filter. (b) Effective potential for a Lorentzian
low-pass filter. Reprinted with permission from Gebhard et al., Phys. Rev. B  107, 064206 (2023).
Copyright (2023) by the American Physical Society.}
 \label{fig:P1_fig6}
\end{figure}

Halperin and Lax~\cite{Halperin1966} suggested instead to use for LF the square of the wave function $\Gamma(x) = \psi^2(x)$. The shape of the wave functions $\psi(x)$ for the low-energy states was determined by the optimal-fluctuation-approach that yields the filtering function

\begin{equation}
  \Gamma^{\rm HL}(x)
  =\frac{1}{2 \ell_{\rm  HL}} \frac{1}{\cosh^2(x/\ell_{\rm HL})} \; ,
  \label{eq:globalHLfilter}
    \end{equation}
where the characteristic length $\ell_{\rm  HL}$ should depend on the state energy~\cite{Halperin1966, Baranovskii1978}. Remarkably, it appears that a universal, energy-independent value for $\ell_{\rm  HL}$, can be introduced~\cite{Gebhard2023}, $\ell_{\rm  HL} = 0.76 \, \ell_0$
as evidenced in Fig.~\ref{fig:P1_fig6}a, where the effective potential yielded by the filtering function given by Eq.~(\ref{eq:globalHLfilter}), 
is compared with the positions and energies of the eigenstates. The eigenstates for the given realization of disorder potential $V(x)$ were obtained via a straightforward solution of the Schr\"{o}dinger equation. In Fig.~\ref{fig:P1_fig6}, the 30 eigenstates with the lowest energies are depicted by red points.
The excellent agreement between the local minima of the effective potential (shown by the solid green line) and the positions and energies of the exactly calculated eigenstates justifies the filter function given by Eq.~(\ref{eq:globalHLfilter}).

In Fig.~\ref{fig:P1_fig6}b, such a comparison is illustrated for the case of a Lorentzian filter, determined by Eq.~(\ref{eq:convolutionLorentz}) with $\ell_{\rm{L}} = 0.27 \, \ell_0$ chosen to mimic the LLT result~\cite{Gebhard2023}. Evidently, the choice of a Lorentzian filter function is not satisfying. The number of the local minima in the effective potential $W(x)$ (shown by the solid blue line) is significantly larger than the true number of the exactly calculated eigenstates. This happens because the Lorentzian function given by Eq.~(\ref{eq:convolutionLorentz}) is not smooth at $x=0$. The cusp at $x=0$ filters too many extrema from the real disorder potential $V(x)$, preventing the identification of true localized electron states by searching the minima of the effective potential $W(x)$. The filter function suggested by Halperin and Lax~\cite{Halperin1966} does not possess such a deficiency.

Not only the energies and spatial positions of localized states in disorder potential $V(\r)$, discussed above, are of interest for the theory. In fact, the key quantity for the optoelectronic properties of disordered semiconductors is the space- and temperature-dependent electron distribution $n(\r,T)$. Below we extend the LF approach to calculate $n(\r,T)$. For that purpose, we introduce the temperature $T$ into the filter function and, concomitantly, into the definition of the effective potential $W(\r,T)$ that yields $n(\r,T)$.


\subsubsection{Universal filter function to determine $n(\r,T)$}
\label{sec:Univ_filter}


The $T$-dependent spatial distribution of electron density $n(\r,T)$ is related to the quasi-classical effective potential $W(\r,T)$ as
\begin{equation}
  n(\r,T) = N_c \exp\left[ \frac{\mu - W(\r,T)}{k_{\rm B} T} \right]\;,
  \label{eq:n-via-W}
\end{equation}
where $N_c$ is the effective density of states in the conduction band and $\mu$ is the chemical potential. This equation serves as the definition of the quasi-classical effective potential $W(\r,T)$. This effective potential is smooth in comparison to the initial disorder potential $V(\r)$ because $W(\r,T)$ is derived from the electron density $n(\r,T)$, which has the spatial scale of the electron wave functions, i.e., is broader than the scale of the short-range fluctuations of $V(\r)$. Let us, therefore, obtain $W(\r,T)$ by subjecting $V(\r)$ to the action of a universal low-pass filter (ULF).



The key question is how to find out the appropriate $T$-dependent filter function $\Gamma(\r,T)$ that can be used to extract the shape of the effective potential $W(\r,T)$  for a given realization of the white-noise potential $V(r)$. One can show that the Fourier image of this function has the shape~\cite{Nenashev2023}
 \begin{equation}
  \hat\Gamma(k) = \frac{\sqrt\pi}{\lambda k} \, e^{-\lambda^2 k^2/4} \, \text{erfi}(\lambda k/2) \; ,
  \label{eq:hat-Gamma-uni}
\end{equation}
where erfi is the imaginary error function, and $\lambda = \hbar/\sqrt{2mk_{\rm B} T}$.

%
%

\begin{figure}
  \includegraphics[width=9cm]{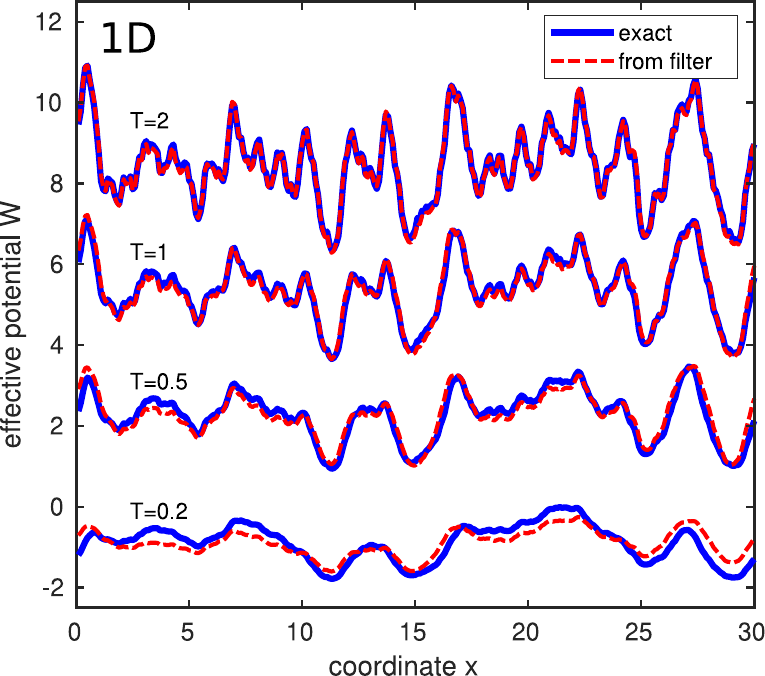}
  \caption{Comparison between the exact effective potential
(solid blue lines) and the filtered potential,  (dashed red
 lines) for a one-dimensional sample with while-noise potential. Reprinted with permission from Nenashev et al., Phys. Rev.
B 107, 064207 (2023). Copyright (2023) by the American Physical Society.}
 \label{fig:P2_fig9}
\end{figure}

\begin{figure}
  \includegraphics[width=9cm]{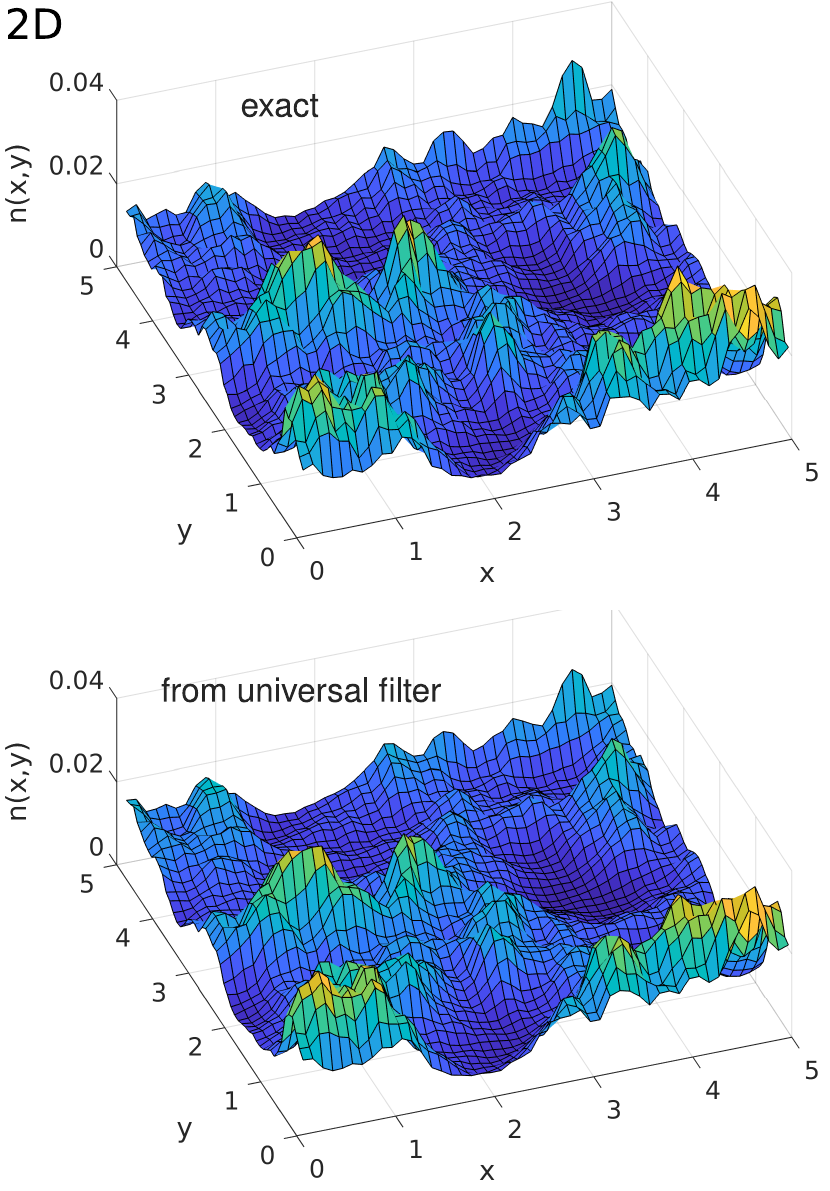}
  \caption{Comparison between the exact electron density
$n(x, y, T)$ (upper part) and that obtained by using the universal filter
function (lower part) in a two-dimensional white-noise potential. Reprinted with permission from Nenashev et al., Phys. Rev.
B 107, 064207 (2023). Copyright (2023) by the American Physical Society.}
 \label{fig:P2_fig13}
\end{figure}

In order to reveal the electron density distribution $n(\r,T)$ for a given realization of the white-noise potential $V(\r)$, one should first calculate the Fourier image $\hat{V}(\k)$ of $V(\r)$ using a fast-Fourier-transform (FFT). This function $\hat{V}(\k)$ should be then multiplied by $\hat{\Gamma}(|\k|)$,
given by Eq.~(\ref{eq:hat-Gamma-uni}). The inverse Fourier transform of the product  $\hat{V}(\k) \hat{\Gamma}(|\k|)$ by the FFT yields the effective potential $W(\r,T)$ because the inverse Fourier transform converts a product into a convolution~\cite{Nenashev2023}. Inserting $W(\r,T)$ into Eq.~(\ref{eq:n-via-W}) gives the electron density $n(\r,T)$. In Fig.~\ref{fig:P2_fig9} and Fig.~\ref{fig:P2_fig13}, we compare the results for $W(\r,T)$ of the above procedure with the effective potentials obtained via Eq.~(\ref{eq:n-via-W}) from the electron density $n(\r,T)$ calculated using the exact solution of Schr\"{o}dinger equation in one and two dimensions, respectively. Coordinate $x$ and temperature $T$ in the figures are measured in units given by Eqs.~(\ref{eq:length-unit}) and (\ref{eq:energy-unit}).


The data in Figs.~\ref{fig:P2_fig9} and \ref{fig:P2_fig13} demonstrate the high accuracy of the approach based on the $T$-dependent low-pass filter. Notably, the FFT operation used in this approach does not need a considerable computer time, in contrast to the exact calculations on the basis of Schr\"{o}dinger equation.

\subsection{Random-wave-functions (RWF) approach to calculate $n(\r,T)$ }
\label{sec:rwf_approach}

\subsubsection{Background}
\label{sec:background_rwf}

The idea of the RWF approach resembles the one suggested recently by Lu and Steinerberger~\cite{Lu2018} to search for the low-lying eigenfunctions of various linear operators. An iterative application of the operator leads to the increasing contributions of the low-energy regions to the state vector~\cite{Lu2018}. A similar approach has been suggested by Krajewski and Parrinello~\cite{Krajewski2005} for the calculation of the thermodynamic potential.

Let us consider the action of the operator $\hat{h}=\exp[-\hat H/(2k_B T)]$ on an arbitrarily chosen wave function. The goal is to model the equilibrium distribution of electrons, which is described by the Boltzmann statistics in the nondegenerate case considered here. In Boltzmann statistics, states with energy $\varepsilon$ contribute to the distribution with the probability proportional to $\exp[-\varepsilon/(k_B T)]$. The wave function is the probability amplitude, which explains the factor $1/2$ in the operator $\hat{h}$. The wave function is always a linear combination of eigenfunctions that correspond to different energies. The action of the operator $\hat{h}$ suppresses the contributions of high-energy eigenfunctions in favor of the contributions of low-energy eigenfunctions. By the application of the operator $\hat{h}$ to a collection of the random wave functions, the average contributions of eigenfunctions corresponding to different energies approach their distribution in thermal equilibrium. The averaging here is performed over the set of the random wave functions. Physically, this procedure corresponds to the averaging of the electron density $n(\r,T)$. The question arises on how to numerically subject a wave function to the action of the operator $\hat{h}=\exp[-\hat H/(2k_B T)]$. It can be done by recursively applying the Hamiltonian $\hat H$ to the wave function:
\begin{equation}
  \hat{h}=e^{-\hat H/2k_BT} \approx (1 - \alpha\hat H)^M \; ,
  \label{eq:exp-H-via-iterations}
\end{equation}
with a natural number~\cite{Nenashev2023} $M \approx 1/(2\alpha k_BT)$
and a small parameter $\alpha$. A simple analysis justifies the choice~\cite{Nenashev2023} $\alpha = 1.5/\epsilon_{\text{max}}$,
where $\epsilon_{\text{max}}$ is the estimate for the upper boundary of the energy distribution. In the case of a regular grid with the lattice constant $a$, $\epsilon_{\text{max}} = \hbar^2/(ma^2)$. Below we describe how to realize this idea technically.

\subsubsection{The RWF algorithm}
\label{subsec:synthesis}

Let us consider the RWF algorithm on a spatial lattice with the volume $\Delta V$ per lattice site.  The value of the random wave function $\psi$ on each lattice site is chosen independently as a random number extracted from a normal distribution with the average value zero and variance $1/\Delta V$.
The following transformation of the wave function $\psi$,
\begin{equation}
  \psi \rightarrow  \psi - \alpha \hat H \psi \,
        \label{eq:step2}
\end{equation}
is applied $M$ times. Then an estimate of the reduced electron density $\tilde{n}_{\rm R}(\r,T)$ is
\begin{equation}
  \tilde n_{\rm R}(\r,T) = 2 \, |\psi(\r)|^2 \, .
  \label{eq:step3}
\end{equation}

The calculation of $\tilde n_{\rm R}(\r,T)$ is carried out for a large number~$N_{\rm R}$ of
realizations~${\rm R}$ of the random wave function $\psi(\r)$.
Then, the electron density $n(\r,T)$ is the arithmetic mean
of the functions $\tilde{n}_{\rm R}(\r)$ obtained for different realizations R,
multiplied by a chemical-potential-related factor $e^{\mu/(k_BT)}$,
\begin{equation}
  n(\r,T) = e^{\mu/(k_BT)} \langle \tilde{n}_{\rm R} (\r,T) \rangle_{\rm R} \; .
  \label{eq:n-from-algorithm}
\end{equation}

\begin{figure}
  \includegraphics[width=9cm]{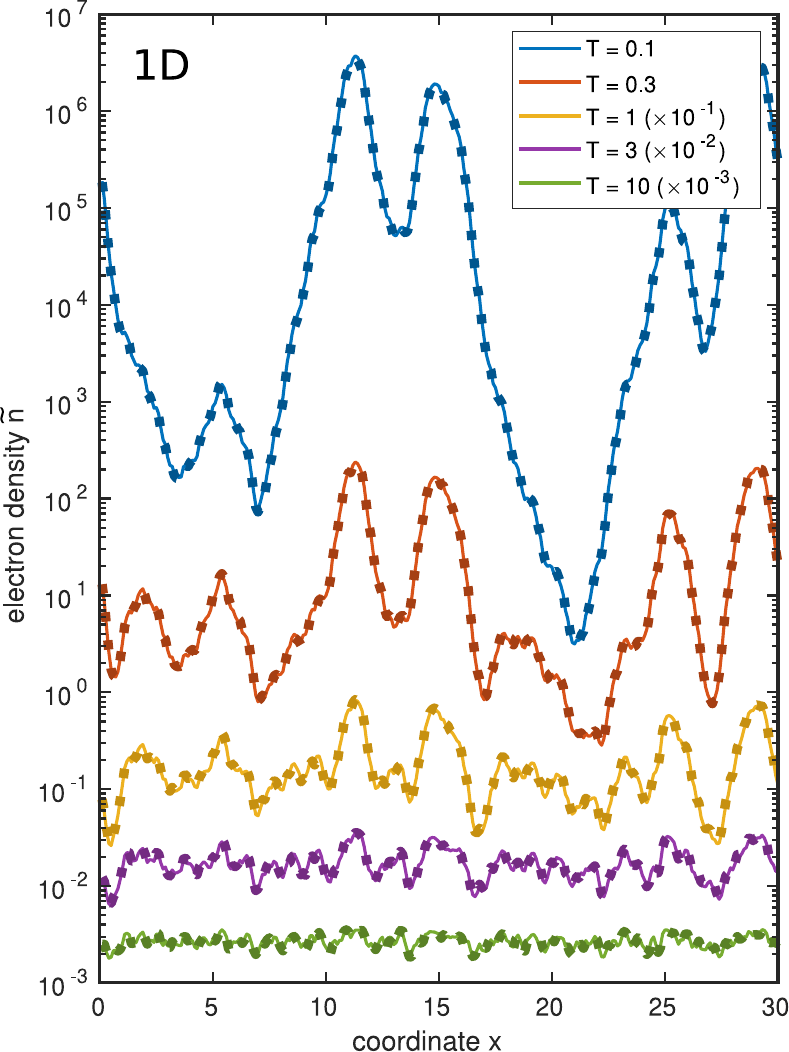}
  \caption{Comparison between exact (solid lines)
          and approximate (symbols) reduced electron density
          $\tilde{n}(x,T) = e^{-\mu/(k_BT)} n(x,T)$
          in a one-dimensional white-noise potential
          at different temperatures~$T$
          and $N_{\rm R} = 1000$ iterations.
          For clarity, the lowest three curves are scaled
          down by a multiplication by $10^{-3}$, $10^{-2}$ and $10^{-1}$,
          as indicated in the legend. Reprinted with permission from Nenashev et al., Phys. Rev.
B 107, 064207 (2023). Copyright (2023) by the American Physical Society.}
 \label{fig:P2_fig3}
\end{figure}
\begin{figure}
  \includegraphics[width=9cm]{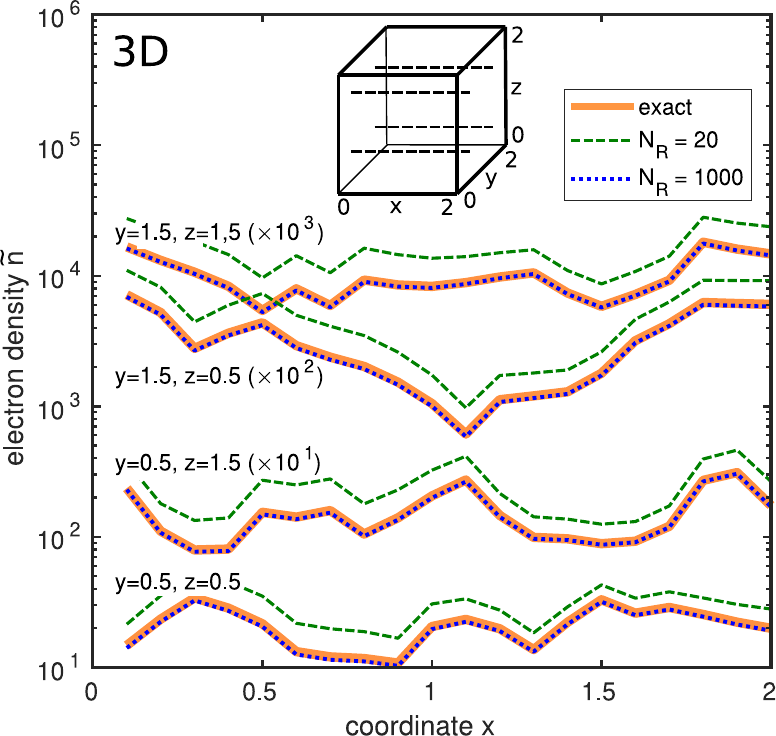}
  \caption{Reduced electron density $\tilde{n}(x,y,z,T)$
          in a three-dimensional sample with white-noise potential.
          Compared are
          the exact density (solid orange line)
          with that obtained by the RWF algorithm
          (dashed green line for $N_{\rm R} = 20$ and dotted blue line for
          $N_{\rm R} = 1000$).
          Profiles along the $x$-axis with different values of
          coordinates $y$ and $z$ are shown (see inset).
          Sample size is $2\times2\times2$ dimensionless units,
          the discretization grid parameter is $a=0.1$,
          the temperature is $T=T_0$.
          Periodic boundary conditions apply.
          For clarity, profiles are multiplied by
          different coefficients,
          as indicated in the plot. Reprinted with permission from Nenashev et al., Phys. Rev.
B 107, 064207 (2023). Copyright (2023) by the American Physical Society.}
 \label{fig:P2_fig5}
\end{figure}

The larger the number of realizations $N_{\rm R}$,
the more accurate is the calculated electron density $n(\r,T)$.

In Fig.~\ref{fig:P2_fig3} and Fig.~\ref{fig:P2_fig5} the reduced electron densities, obtained via the RWF algorithm are compared with the exact ones calculated using the solution of the Schr\"{o}dinger equation for one and three dimensions, respectively. Evidently, the RWF algorithm with 1000 iterations accurately yields the electron density in a wide range of temperatures.

\section{Discussion}
\label{sec:discussion}

In this work, we introduce four theoretical tools to get access to the space- and temperature-dependent electron density $n(\r,T)$ in disordered media with a random potential $V(\r)$, avoiding the time-consuming numerical solution of the Schr\"{o}dinger equation.

For the case of degenerate conditions controlled by Fermi statistics, the Hamiltonian is converted into a matrix, which, being subjected to several multiplications with itself succeeded by inverting the outcome, yields the distribution $n(\r,T)$. The other possible technique for the case of Fermi statistics replaces the operation of matrix inversion by solving a system of linear equations controlled by the matrix generated from the Hamiltonian.

For non-degenerate conditions with Boltzmann statistics, the universal low-pass filter (ULF) approach and the random-wave-function
(RWF) algorithm are suggested for approximate calculations of $n(\r,T)$.
Both methods require far less computational resources than
the complete solution of the Schr\"odinger equation.

The ULF approach employs the temperature-de\-pen\-dent effective potential $W(\r,T)$.
This technique is based on the Fast Fourier Transformation, which does not impose any demands on computational resources, such as processor time and memory. Therefore, it can be applied to mesoscopically large three-dimensional
disordered systems.

Being superior to the widely used approximate methods, the
RWF is computationally more costly than the ULF approach, if mesoscopically large three-dimensional systems at low temperatures are addressed. However, the accuracy of calculations based on the RWF algorithm can be unlimitedly improved by increasing the number of the RWF realizations.


\newpage
\section*{Appendix}
\appendix

\section{Why do Methods 1 and 2 work}
\label{sec:why}

To understand why Method~1 works, let us apply to the Hamiltonian $\hat H$ a unitary transformation that diagonalizes it. This transformation also diagonalizes all the matrices that appear in Method~1. The diagonal elements of matrices $\hat H$, $\hat A_0$, $\hat A_N$ and $\hat B$ are, in accord to Eqs.~(\ref{eq:make_A0}) -- (\ref{eq:make-B}),
\begin{align}
	H_{aa} &= \ve_a \, , \\
	A_{0,aa} &= \frac{\ve_a - \ve_0}{\ve_f - \ve_0} \, , \\
	A_{N,aa} &= \left( \frac{\ve_a - \ve_0}{\ve_f - \ve_0} \right)^{2^N} \, , \\
	B_{aa} &= \frac{1}{ \left( \frac{\ve_a - \ve_0}{\ve_f - \ve_0} \right)^{2^N} + 1} \, . \label{eq:B-diag-elements}
\end{align}
Let us consider those values of $\ve_a$ that are close to the Fermi energy $\ve_f$. For them,
\begin{equation}\label{eq:AN-approx}
	\left( \frac{\ve_a - \ve_0}{\ve_f - \ve_0} \right)^{2^N} = \left( \frac{\ve_a - \ve_f}{\ve_f - \ve_0} + 1 \right)^{2^N} \approx \exp \left( 2^N \frac{\ve_a - \ve_f}{\ve_f - \ve_0} \right) .
\end{equation}
Taking Eq.~(\ref{eq:parameters-and-T}) into account, one can rewrite this estimate as
\begin{equation}\label{eq:AN-approx-via-kT}
	\left( \frac{\ve_a - \ve_0}{\ve_f - \ve_0} \right)^{2^N} \approx \exp \left( \pm \, \frac{\ve_a - \ve_f}{k_BT} \right) ,
\end{equation}
where sign ``$+$'' is chosen if $\ve_0 < \ve_f$, and sign ``$-$'' otherwise. Substitution of Eq.~(\ref{eq:AN-approx-via-kT}) into Eq.~(\ref{eq:B-diag-elements}) and comparison to expression~(\ref{eq:f-definition}) for the Fermi function $f(\ve)$ provides the following result:
\begin{equation}\label{eq:B-diag-elements1}
	B_{aa} \approx f(\ve_a) \quad \mathrm{if} \quad \ve_0 < \ve_f \, ,
\end{equation}
\begin{equation}\label{eq:B-diag-elements2}
	1 - B_{aa} \approx f(\ve_a) \quad \mathrm{if} \quad \ve_0 > \ve_f \, ,
\end{equation}
which is valid after the unitary transformation that diagonalizes the Hamiltonian. If we undo this transformation, Eq.~(\ref{eq:B-diag-elements1}) acquires the form
\begin{equation}\label{eq:B-approx1}
	B_{ij} \approx \sum_{a=1}^L f(\ve_a) \psi_{a,i} \psi_{a,j}^* \quad \mathrm{if} \quad \ve_0 < \ve_f \, ,
\end{equation}
that is, the diagonal matrix elements of the matrix $\hat B$ are equal to
\begin{equation}\label{eq:B-approx2}
	B_{ii} \approx \sum_{a=1}^L f(\ve_a) |\psi_{a,i}|^2 \quad \mathrm{if} \quad \ve_0 < \ve_f \, ,
\end{equation}
or, due to Eq.~(\ref{eq:n-discrete}),
\begin{equation}\label{eq:B-approx3}
	B_{ii} \approx \frac{\Delta V}{2} \, n_i \quad \mathrm{if} \quad \ve_0 < \ve_f \, ,
\end{equation}
that justifies Eq.~(\ref{eq:n-by-inversion-1}) of Method~1.

Similarly, from Eq.~(\ref{eq:B-diag-elements2}) one can get after undoing the unitary transformation:
\begin{equation}\label{eq:B-approx4}
	1 - B_{ii} \approx \frac{\Delta V}{2} \, n_i \quad \mathrm{if} \quad \ve_0 > \ve_f \, ,
\end{equation}
which justifies Eq.~(\ref{eq:n-by-inversion-2}) of Method~1.

Now let us consider Method~2. For simplicity, we restrict ourselves to the case $\ve_0 < \ve_f$. The opposite case is completely analogous.

\begin{figure}
\includegraphics[width=3cm]{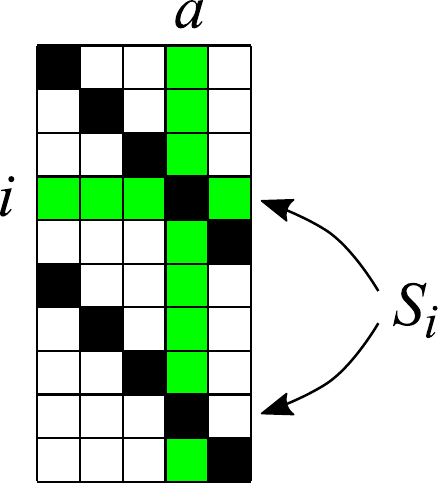}
\caption{Illustration of set $S_i$ for a given row index $i$ of matrix $\hat U$. White and black squares represent zeros and ones, respectively. Row $i$ and column $a$ of the matrix are highlighted in green.}
\label{fig:U2}
\end{figure}

The matrix $\hat X$ can be expressed as $\hat X = \hat B \hat U$ according to Eqs.~(\ref{eq:make-B}) and~(\ref{eq:linear-equations}). Hence, the sum in the right-hand side of Eq.~(\ref{eq:n-by-linear-equations-1}) can be rewritten as
\begin{equation}\label{eq:sum-rewritten1}
	\sum_{a=1}^{N_C} X_{ia} U_{ia} = \sum_{a=1}^{N_C} \sum_{j=1}^{L} B_{ij} U_{ja} U_{ia} \, .
\end{equation}
For a given $i$, there is only one value of $a$ such that $U_{ia}=1$, according to construction of the matrix $\hat U$ (see Section~\ref{sec:linear-equations}). For this value of $a$, there are only few values of $j$ such that $U_{ja}=1$, as illustrated in Fig~\ref{fig:U2}. Let us denote the set of these values of $j$ as $S_i$. One can therefore rewrite Eq.~(\ref{eq:sum-rewritten1}) as
\begin{equation}\label{eq:sum-rewritten2}
	\sum_{a=1}^{N_C} X_{ia} U_{ia} = \sum_{j \in S_i} B_{ij} \, .
\end{equation}
Note that $i \in S_i$ since $U_{ia}=1$. Therefore the sum in the right-hand side of Eq.~(\ref{eq:sum-rewritten2}) contains the term $B_{ii}$. 

Let us argue now that, at large enough $N_C$, all other terms $B_{ij}$ in this sum with $j \neq i$ are negligible.
One can see from Eq.~(\ref{eq:B-approx1}) that the matrix $\hat B$ is close to the projector $\hat P$ to the set of eigenstates with energies below the Fermi energy. Typically, especially in the situation when the filled states are separated by a band gap from the empty ones, the matrix elements $P_{ij}$ of this projector decrease with increasing the distance between nodes $i$ and $j$, and become negligible at some distance. Hence, the matrix elements $B_{ij}$ in the right-hand side of Eq.~(\ref{eq:sum-rewritten2}) becomes negligible if nodes $i$ and $j$ are far away from each other, i.~e., if $i \neq j$. The only non-negligible term is $B_{ii}$, and therefore,
\begin{equation}\label{eq:sum-rewritten3}
	\sum_{a=1}^{N_C} X_{ia} U_{ia} \approx  B_{ii} \, .
\end{equation}
Substituting there $B_{ii}$ from Eq.~(\ref{eq:B-approx3}), one arrives at Eq.~(\ref{eq:n-by-linear-equations-1}) of Method~2.

Similarly, the combination of Eqs.~(\ref{eq:B-approx4}) and~(\ref{eq:sum-rewritten3}) justifies  Eq.~(\ref{eq:n-by-linear-equations-2}) of Method~2 in the case $\ve_0 > \ve_f$.


\section{Choice of parameters $\ve_0$, $N$ and $N_C$}
\label{sec:choice}

There are two restrictions on the choice of parameters $\ve_0$ and $N$ and, hence, on the temperature value $T$ that can be achieved with Methods~1 and~2. First, the approximate equalities~(\ref{eq:B-diag-elements1}) and~(\ref{eq:B-diag-elements2}) must hold in the whole range of electron energies. And second, the matrix $(\hat A_N + \hat I)$ must not be ill-conditioned.

To consider the first restriction, let us define the function $\tilde f(\ve)$
\begin{equation}\label{eq:f-tilde-definition-1}
	\tilde f(\ve) = \frac{1}{ \left( \frac{\ve - \ve_0}{\ve_f - \ve_0} \right)^{2^N} + 1} \quad \mathrm{if} \quad \ve_0 < \ve_f \, ,
\end{equation}
\begin{equation}\label{eq:f-tilde-definition-2}
	\tilde f(\ve) = 1 - \frac{1}{ \left( \frac{\ve - \ve_0}{\ve_f - \ve_0} \right)^{2^N} + 1} \quad \mathrm{if} \quad \ve_0 > \ve_f \, .
\end{equation}
Methods~1 and~2 are based on the fact that this function is close to the Fermi function $f(\ve)$ in a vicinity of the Fermi energy $\ve_f$, see Eqs.~(\ref{eq:B-diag-elements}), (\ref{eq:B-diag-elements1}) and~(\ref{eq:B-diag-elements2}). Let us now consider the behavior of the function $\tilde f(\ve)$ in a broad range of energies $\ve$.

\begin{figure}
\includegraphics[width=10cm]{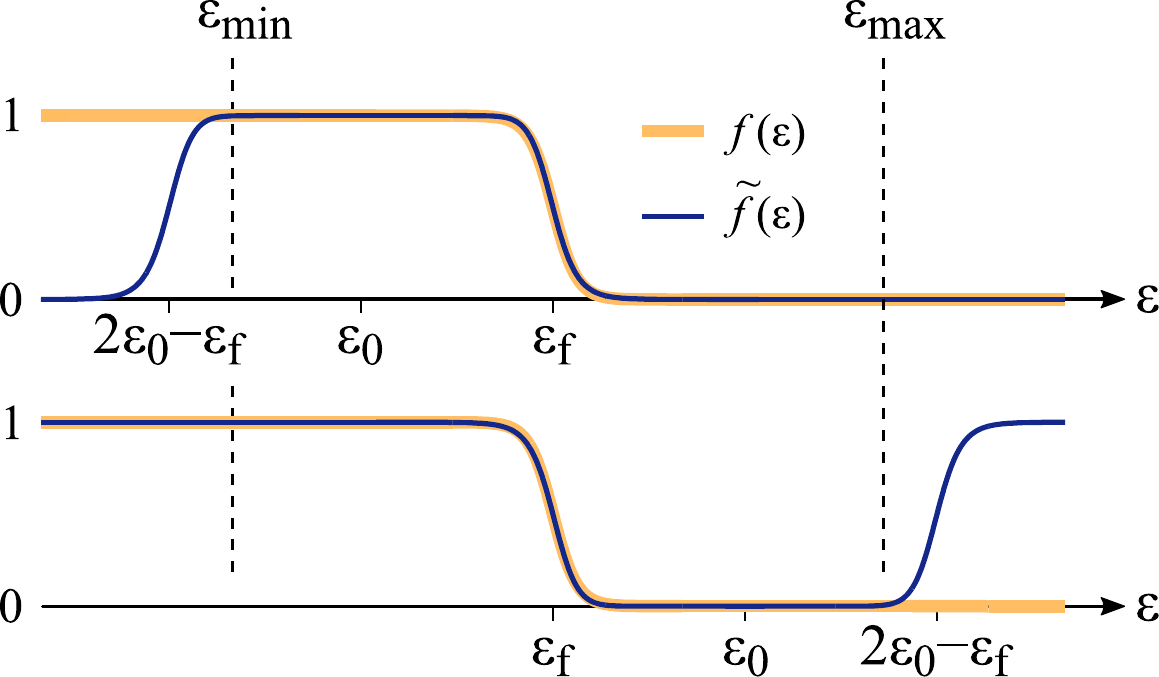}
\caption{Comparison of the Fermi function $f(\ve)$ and function $\tilde f(\ve)$. Above: the case of $\ve_0 < \ve_f$, below: the case of $\ve_0 > \ve_f$.}
\label{fig:f-tilde}
\end{figure}

As an example, Fig.~\ref{fig:f-tilde} shows $\tilde f(\ve)$ along with the Fermi function $f(\ve)$ for $N=4$ and two choices of the ``reference energy'' $\ve_0$: $\ve_0 < \ve_f$ (upper panel) and $\ve_0 > \ve_f$ (lower panel). In both cases, there is a discrepancy between the functions $f(\ve)$ and $\tilde f(\ve)$ below the energy $2\ve_0-\ve_f$ in the first case, and above the energy $2\ve_0-\ve_f$ in the second case. Electron energy levels must not fall into these areas of discrepancy, otherwise the contributions of these levels into the electron density would not be accounted correctly in Methods~1 and~2. Therefore, in the case of $\ve_0 < \ve_f$, the lowest electron energy $\ve_{\mathrm{min}}$ must be larger than $2\ve_0-\ve_f$. Similarly, in the case of $\ve_0 > \ve_f$, the highest electron energy $\ve_{\mathrm{max}}$ must be smaller than $2\ve_0-\ve_f$. These conditions can be rewritten as restrictions to the ``reference energy'' $\ve_0$:
\begin{equation}\label{eq:restriction-eps0}
	\ve_0 < \frac{\ve_f + \ve_{\mathrm{min}}}{2} \quad \mathrm{or} \quad \ve_0 > \frac{\ve_f + \ve_{\mathrm{max}}}{2} \, .
\end{equation}

Now let us consider the second restriction. The inversion of the matrix $(\hat A_N + \hat I)$ in Method~1, or solving a system of linear equations expressed by this matrix in Method~2, is possible when this matrix is not ill-conditioned. This means that the condition number, a ratio of the largest and the smallest eigenvalues of the matrix, is less than the maximal value of the order of $10^{15}$ (a number that corresponds to the accuracy of representation of real numbers in the computer memory). The minimal eigenvalue of matrix $(\hat A_N + \hat I)$ is close to unity, and corresponds to energy levels near to $\ve_0$. The largest eigenvalue corresponds to the energy level farthest from $\ve_0$, i.~e., either to $\ve_{\mathrm{min}}$ or to $\ve_{\mathrm{max}}$, and is approximately the largest of two values $\left[ (\ve_{\mathrm{min}} - \ve_0)/(\ve_f - \ve_0) \right]^{2^N}$ and $\left[ (\ve_{\mathrm{max}} - \ve_0)/(\ve_f - \ve_0) \right]^{2^N}$. Hence, the parameters $\ve_0$ and $N$ have to obey the following restrictions:
\begin{equation}\label{eq:restriction-ill-conditioned}
	\left( \frac{\ve_{\mathrm{min}} - \ve_0}{\ve_f - \ve_0} \right)^{2^N} < 10^{15} \quad \mathrm{and} \quad
	\left( \frac{\ve_{\mathrm{max}} - \ve_0}{\ve_f - \ve_0} \right)^{2^N} < 10^{15} \, .
\end{equation}

The choice of parameters $\ve_0$ and $N$ is based on equations~(\ref{eq:restriction-eps0}) and~(\ref{eq:restriction-ill-conditioned}). We consider two different options: (i) when the temperature $T$ is given, and (ii) when the goal is to obtain the sharpest possible boundary between filled and empty states.

In the first option, one should try the natural numbers in ascending order as values of $N$. For each such number $N$, one finds $\ve_0$ according to Eq.~(\ref{eq:parameters-and-T}) and check whether conditions~(\ref{eq:restriction-eps0}) and~(\ref{eq:restriction-ill-conditioned}) are fulfilled. The smallest suitable value of $N$ is the best choice. Indeed, the smaller $N$ is, the more sparse matrix $\left( \hat A_N + \hat I \right)$ is and, consequently, the faster is matrix inversion in Method~1 or solution of linear equations in Method~2.

In the second option, the largest possible $N$ is desirable, in order to minimize the temperature $T$ according to Eq.~(\ref{eq:parameters-and-T}). To achieve the largest $T$, one has to choose the value $\ve_0$ that maximizes the denominator $|\ve_f - \ve_0|$ in Eq.~(\ref{eq:restriction-ill-conditioned}):
\begin{equation}\label{eq:choice-of-eps0}
	\ve_0 \approx \frac{\ve_f + \ve_{\mathrm{min}}}{2} \quad \mathrm{or} \quad \ve_0 \approx \frac{\ve_f + \ve_{\mathrm{max}}}{2} \, ,
\end{equation}
and then one needs to choose the maximal number $N$ that obeys the restrictions~(\ref{eq:restriction-ill-conditioned}),
\begin{equation}\label{eq:choice-of-N-1}
	N \approx \log_2 \left( \frac{15}{ \log_{10}\left| \frac{\ve_{\mathrm{max}} - \ve_0}{\ve_f - \ve_0} \right| } \right)
\end{equation}
or
\begin{equation}\label{eq:choice-of-N-2}
	N \approx \log_2 \left( \frac{15}{ \log_{10}\left| \frac{\ve_{\mathrm{min}} - \ve_0}{\ve_f - \ve_0} \right| } \right)
\end{equation}
for the first and the second choice of $\ve_0$ in Eq.~(\ref{eq:choice-of-eps0}), correspondingly.

Finally, let us consider the choice of the parameter $N_C$ in Method~2. The accuracy of Method~2 improves with $N_C$. However, larger $N_C$ give rise to a longer computation time due to the increase of the size of the matrix $\hat U$ in Eq.~(\ref{eq:linear-equations}). Hence, there is a trade-off between the accuracy and the efficiency. The best value of parameter $N_C$ can be estimated as
\begin{equation}\label{eq:choice-of-NC}
	N_C \simeq (\ell/a)^d ,
\end{equation}
where $d$ is dimensionality of the space, $a$ is the distance between neighboring sites in the lattice, and $\ell$ is a characteristic length of decay of the matrix element $B_{ij}$ with increasing the distance between sites $i$ and $j$.

\begin{acknowledgement}

S.D.B. and K.M. acknowledge financial support by the Deutsche
Forschungsgemeinschaft (Research Training Group ``TIDE'', RTG2591)
as well as by the key profile area ``Quantum Matter and Materials
(QM2)'' at the University of Cologne. K.M. further acknowledges support
by the DFG through the project ASTRAL (ME1246-42).

\end{acknowledgement}

\bibliography{LL_filter}

\end{document}